# ANALYSIS OF ULTRA-SHORT BUNCHES IN FREE-ELECTRON LASERS

L.T. Campbell,[1,2] H.P. Freund,[3] J.R. Henderson,[1,2] B.W.J. McNeil,[1,4] P. Traczykowski[1,2,4] and P.J.M. van der Slot[5]

[1]University of Strathclyde (SUPA), Glasgow G4 0NG, United Kingdom
[2]ASTeC, STFC Daresbury Laboratory and Cockcroft Institute, Warrington WA4 4AD, United Kingdom
[3]University of New Mexico, Albuquerque, New Mexico, 87131 United States of America
[4]Cockcroft Institute, Warrington WA4 4 AD, United Kingdom
[5]Laser Physics and Nonlinear Optics, Mesa+ Institute for Nanotechnology, University of Twente, Enschede, the Netherlands

Free-electron lasers (FELs) operate at wavelengths from millimeter waves through hard x-rays. At x-ray wavelengths, FELs typically rely on self-amplified spontaneous emission (SASE). Typical SASE emission contains multiple temporal "spikes" which limit the longitudinal coherence of the optical output; hence, alternate schemes that improve on the longitudinal coherence of the SASE emission are of interest. In this paper, we consider electron bunches that are shorter than the SASE spike separation. In such cases, the spontaneously generated radiation consists of a single optical pulse with better longitudinal coherence than is found in typical SASE FELs. To investigate this regime, we use two FEL simulation codes. One (MINERVA) uses the slowly-varying envelope approximation (SVEA) which breaks down for extremely short pulses. The second (PUFFIN) is a particle-in-cell (PiC) simulation code that is considered to be a more complete model of the underlying physics and which is able to simulate very short pulses. We first anchor these codes by showing that there is substantial agreement between the codes in simulation of the SPARC SASE FEL experiment at ENEA Frascati. We then compare the two codes for simulations using electron bunch lengths that are shorter than the SASE slice separation. The comparisons between the two codes for short bunch simulations elucidate the limitations of the SVEA in this regime but indicate that the SVEA can treat short bunches that are comparable to the cooperation length.

## I. INTRODUCTION

While free-electron lasers (FELs) have been intensively studied since the 1970s, new developments and concepts keep the field fresh. Intensive work is ongoing into new FEL-based light sources that probe ever shorter wavelengths with a variety of configurations. There presently exists a large variety of FELs ranging from long-wavelength oscillators using partial wave guiding to ultraviolet and hard x-ray FELs that are either seeded or start from noise (*i.e.*, self-amplified spontaneous emission or SASE). As new FEL light sources come on-line, interest will grow in shorter pulses, new spectral ranges and higher photon fluxes. The increasing activity in the design and construction of FEL light sources is associated with increasing simulation activity to design, optimize, and characterize these FELs and test novel device concepts. In particular, concepts that improve on the longitudinal coherence of SASE FELs can have important practical applications.

In this paper, we consider the use of ultra-short electron bunches in SASE FELs. By "ultra-short" we mean electron bunches that are comparable to or shorter than the cooperation length and, hence, are shorter than the SASE spike separation. In such cases, the spontaneously generated emission consists of a single optical pulse which has improved longitudinal coherence with respect to what is produced in a typical spiky SASE FEL [1,2].

Most of the physical/numerical models used to describe the FEL are based on either the slowly-varying envelope approximation (SVEA) or a particle-in-cell (PiC) formulation. These two formulations have different limitations. On the one hand, while the SVEA codes require relatively modest computational resources the formulation breaks down for sufficiently short pulses because they employ an average of Maxwell's equations over the time scale of the resonant wave. On the other hand, the PiC formulation integrates the unaveraged Maxwell's equations and does not make a resonant wave approximation. However, while a PiC model can treat arbitrarily short pulses, they require much larger computational resources than the SVEA codes.

In the SVEA, the optical field is represented by a slowly-varying amplitude and phase with respect to the rapid sinusoidal oscillation of the carrier frequency. The field equations are then averaged over the rapid sinusoidal time scale and, thereby, reduced to equations describing the evolution of the slowly-varying amplitude and phase. Within the context of the SVEA, FEL simulation codes fall into two main categories where the particle trajectories are found by first averaging the trajectories over an undulator period (the so-called KMR approximation), or by the direct integration of the Newton-Lorentz equations. There is a further distinction between the SVEA codes based upon the optical field representation, and codes have been written using either a grid-based field solver or a superposition of optical modes. Simulation codes using the KMR analysis in conjunction with a grid-based field solver include (but are not limited to) GINGER [3], GENESIS [4], and FAST [5]. In contrast, SVEA codes that integrate the Newton-Lorentz equations in conjunction with a Gaussian mode superposition for the optical fields include MEDUSA [6] and MINERVA [7]. One common feature of all the SVEA codes, however, is the way in which time-dependence is treated. The fast time scale average results in a breakdown of the optical pulse and the electron beam into temporal *slices* each of which is one wave period in duration. The optical slices slip ahead of the electron slices at the rate of one wavelength per undulator period. As a result, the SVEA codes integrate each electron and optical slice from $z \rightarrow z + \Delta z$ and then allow the optical slice to slip ahead of the electron slices. Although these codes have been extremely successful in modeling FELs the SVEA breaks down for sufficiently short optical/electron pulse interactions.

In contrast, PiC codes are considered to represent a more fundamental model of FEL physics. A PiC code makes no average over the rapid sinusoidal oscillation and integrates the Newton-Lorentz equations for the particles as well as Maxwell's equations for the fields. As a result, PIC codes are able to simulate short pulse interactions but require substantially more computational resources than SVEA codes and are not so commonly used and have not been as extensively validated against experiments as have the SVEA codes. At the present time, the primary PiC code for FEL simulations is PUFFIN [8].

In this paper we present an analysis of ultra-short bunches in FELs using both the SVEA and PiC numerical formulations and use the MINERVA code and PUFFIN as representations of these approaches. In order to benchmark the two codes, we first compare them with experimental measurements from a longer pulse SASE FEL and then proceed to short pulse simulations. We emphasize that the two codes share no common elements. In particular, the particle loading algorithms used to treat start-up from noise are different. MINERVA uses an adaptation of the algorithm described in [9] while PUFFIN uses an algorithm developed in [10].

The organization of the paper is as follows. To illustrate the differences in the formulations contained within the two codes, a brief description of MINERVA and PUFFIN is given in Sec. II. In order to provide a basis for comparison of the two codes, we first compare them in Sec. III with experimental measurements at the "Sorgente Pulsata ed Amplificata di Radiazione Coerente" (SPARC) experiment which is a SASE FEL located at ENEA Frascati [11]. The two codes are then used to simulate short electron bunches and SASE generation based on the configuration of the SPARC FEL is described in Sec. IV. A summary with conclusions is given in Sec. V.

## II. THE NUMERICAL FORMULATIONS

Maxwell's equations are averaged over a wave period in the SVEA formulation. This means that the fast oscillation frequency is not resolved in MINERVA and that simulations can be performed more rapidly than is possible in the PUFFIN PiC simulation in which the field equations are unaveraged. MINERVA treats steady-state simulation by the simple expedient of including a single temporal *slice* in the simulation. Here, a *slice* refers to a segment of the electron beam that is one wavelength long. Time dependence is treated by including multiple temporal *slices* and allowing the field *slices* to advance relative to the electron *slices* at arbitrary integration intervals. Since the optical field slips ahead of the electrons at the rate of one wavelength per undulator period, if this slippage operation is performed at shorter intervals than the undulator period (which is typically after each of 20 – 30 steps per undulator period), then the field advance is interpolated between adjacent temporal *slices* based on this slippage rate. MINERVA uses a 4$^{th}$ order Runge-Kutta integrator to integrate both the particle and field equations; hence, it is 4$^{th}$ order accurate.

PUFFIN uses a split-step Fourier-RK4 method as described in [8]. Each step is decomposed into a free-space radiation diffraction step in the absence of the electrons, and a step solving the radiation generation from the electron beam concurrently with the electron Lorentz force equations using a standard 4$^{th}$ order Runge-Kutta method. The finite element method is not used, as the electron macroparticles are interpolated directly onto the field mesh cells, removing the need for an external linear solver. PUFFIN does not average over the fast time scale and must resolve oscillations on all time and space scales including the undulator and optical periods. As such, the electron beam is not divided into slices in PUFFIN; rather, the bunch is simulated as a whole. This means that PUFFIN requires a larger computational investment than for either MINERVA or other SVEA codes.

Both MINERVA [7] and PUFFIN [8] describe the particles and fields in three spatial dimensions. Electron trajectories are integrated using the complete Newton-Lorentz force equations using the magnetostatic and electromagnetic fields. No wiggler-averaged-orbit approximation is made in either code; hence, the wiggle-motion in the undulators must be resolved. In practice, this means that MINERVA takes 20 – 30 steps per undulator period. In this, MINERVA is similar to PUFFIN which also must resolve the wiggler-motion. However, PUFFIN must also resolve the optical scales which can require anywhere from 10 – 20 steps/grid cells per optical period/wavelength.

The magnetostatic fields are specified by analytical functions for several undulator models (such as planar, elliptical, or helical representations), quadrupoles, and dipoles. These magnetic elements can be placed in arbitrary sequences for any given transport lines. As such, field configurations can be specified for single or multiple undulator segments with quadrupoles either placed between the undulators or superimposed upon the undulators to create a FODO lattice. Dipole chicanes can also be placed between the undulators to model various optical klystron and/or high-gain harmonic generation (HGHG) configurations.

The electromagnetic field is described by a modal expansion in MINERVA. Gaussian optical modes are used for free-space propagation. The Gauss-Hermite modes are used for simulation of planar undulators, while Gauss-Laguerre modes are used for elliptical or helical undulators. In contrast, PUFFIN uses a grid-based cell algorithm for the electromagnetic fields.

It is important to remark that the formulations and detailed coding in PUFFIN and MINERVA share no common elements. In particular, different particle loading algorithms are used in MINERVA and PUFFIN when simulating SASE.

## III. THE SPARC SASE FEL

We employ the SPARC [11] as the basis for comparison and validation fo the two codes. The best estimate for the experimental parameters of SPARC are summarized in Table 1 There were six undulator modules in the experiment (with a one period up-taper at each undulator entrance and one for the exit down-taper). The quadrupoles formed a strong focusing lattice and were located 0.105 m downstream from the exit of the previous undulator module. Note that the quadrupole orientations were fixed and did not alternate. The electron beam was matched into the undulator/focusing lattice and the resonant radiation wavelength was 491.5 nm.

The six undulators give an overall length of approximately 15 meters; however, this was too short to reach saturation for the given bunch charge. In order to compare the codes in the saturated regime, therefore, w the undulator/FODO lattice was extended to include 11 undulator modules with a total length of approximately 28 meters. As a result, the experimental data is used to anchor the validation of the codes in the start-up and exponential growth regions, while the code results are compared for the initial start-up, exponential growth and deep saturation regimes. In the experiment, the pulse energies were measured in the gaps after each undulator segments by opening the gaps in successive undulators, thereby detuning the FEL interaction, in the further downstream undulators [11].

| **Electron Beam** | |
|---|---|
| Energy | 151.9 MeV |
| Bunch Charge | 450 pC |
| rms Bunch Duration | 2.83 ps |
| $x$-Emittance | 2.5 mm-mrad |
| $y$-Emittance | 2.9 mm-mrad |
| rms Energy Spread | 0.02% |
| rms Size ($x$) | 132 μm |
| $\alpha_x$ | 0.938 |
| rms Size ($y$) | 75 μm |
| $\alpha_y$ | -0.705 |
| **Undulator Modules** | |
| Period | 2.8 cm |
| Length | 77 Periods |
| Amplitude | 7.8796 kG |
| $K_{rms}$ | 1.457 |
| Module Gap | 0.40 m |
| **Quadrupoles** | |
| Length | 5.3 cm |
| Field Gradient | 0.9 kG/cm |

Table 1: Parameters of the SPARC FEL experiment.

The simulated propagation of the beam through the undulator/quadrupole lattice is shown in Fig. 1, where we plot the beam envelope in $x$ (blue, left axis) and $y$ (red, right axis) versus position as determined by MINERVA. The PUFFIN propagation results are similar. Observe that the beam is well-confined over the 28 meters of the extended lattice with an average beam size of approximately 115 microns.

A comparison of the evolution of the pulse energy as found in MINERVA and PUFFIN, and as measured in the experiment, is shown in Fig. 2 where the MINERVA simulation is indicated by the blue line and the PUFFIN simulation is indicated by the green line. Note that this represents an average over 15 runs with different noise seeds in MINERVA and 5 runs with different noise seeds in PUFFIN. The reason for the smaller number of runs with PUFFIN is that each run requires considerably more computer time than needed for MINERVA runs. However, convergence is achieved over simulations using multiple noise seeds to within about 5% for both codes. The measured pulse energies are indicated by the red markers where the error bars indicate the standard deviation over a sequence of shots (data courtesy of L. Giannessi). Observe that the agreement between the two codes, and between the codes and the measured pulse energies, is excellent over the entire range of the experiment.

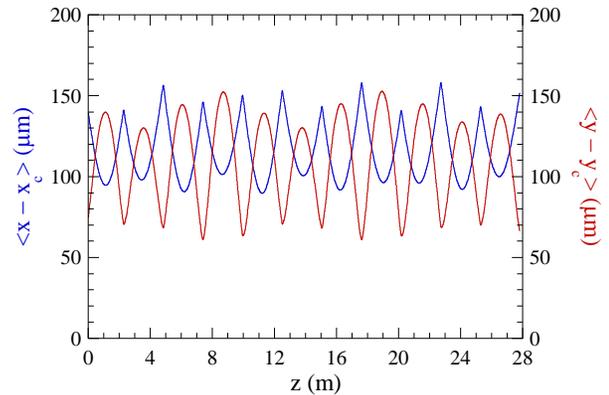

Fig. 1: MINERVA simulation of the beam propagation through the undulator/quadrupole lattice.

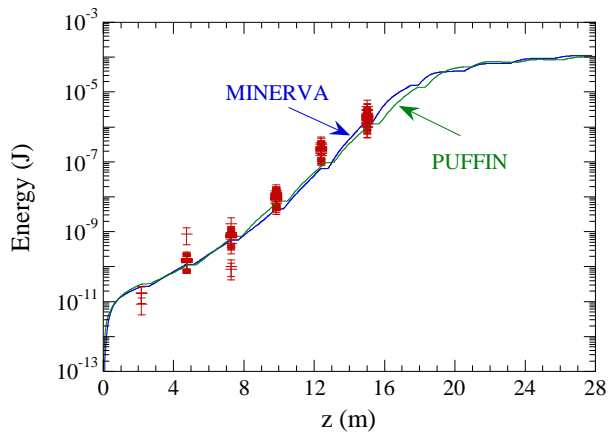

Fig. 2: Comparison of simulation results with PUFFIN and MINERVA and the measured pulse energies versus distance through the undulator (data courtesy of L. Giannessi).

As shown in Fig. 2, exponential growth starts in the second undulator module and that the start-up region is encompassed in the first module. The experimental measurements indicate that the pulse energy after the first undulator falls into the range of approximately $8.4 \times 10^{-12}$ J to $1.74 \times 10^{-11}$ J while MINERVA yields an average pulse energy of $2.52 \times 10^{-11}$ J and PUFFIN yields $3.15 \times 10^{-11}$ J resulting in a discrepancy of only about 24% between the two codes. That the experimental value is somewhat lower than the simulations is to be expected as it is measured at some distance downstream from the first undulator module while the codes evaluate the pulse energy at every location along the undulator line. Hence, the simulation results are in relatively close agreement with the experiment and with each other. This agreement is an important observation since the particle loading algorithms in the two codes share no commonality. Apart from differences that might derive from the parabolic versus Gaussian temporal profiles and the different particle loading algorithms, another source of the difference in the slightly higher start-up noise in PUFFIN is the fact that PUFFIN naturally includes a wider initial spectral range than MINERVA.

The exponential growth region starts in the second undulator module and the two codes are in close agreement with each other and with the experimental measurements out to the end of the sixth undulator. These results are in substantial agreement with the parameterization developed by Ming Xie [12]. Using a $\beta$-function of approximately 2 m, we find that the Pierce parameter $\rho \approx 2.88 \times 10^{-3}$ and that this parameterization predicts a gain length of 0.67 m, and a saturation distance of 18.1 m (including the additional 3.2 m represented by the gaps between undulator modules). This is in reasonable agreement with the simulations which indicate that saturation occurs after between about 18 – 20 meters of undulator/FODO line.

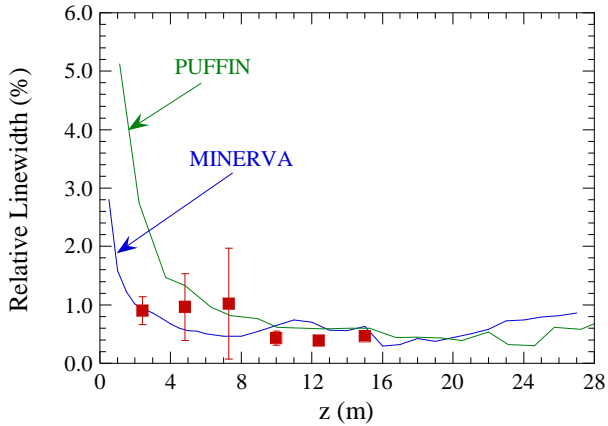

Fig. 3: Comparison of the measured relative linewidth in red (data courtesy of L. Giannessi) with that found in the simulations (blue for MINERVA and green for PUFFIN).

Finally, the predictions of the two codes in the saturation regime after approximately 20 m are also in very good agreement. After 28 m of undulator/FODO lattice, PUFFIN predicts a pulse energy of 104 µJ while MINERVA predicts 111 µJ for a difference of 6.3%.

The larger initial spectral linewidth excited in the start-up region exhibited by PUFFIN is shown more clearly in Fig. 3 which presents a comparison between the evolution of the relative rms linewidth as determined from PUFFIN and MINERVA and by measurement (data courtesy of L. Giannessi). It is clear that a PiC code such as PUFFIN predicts the generation of a wider initial bandwidth of incoherent spontaneous emission. Exponential gain due to the resonant FEL interaction starts in the second undulator module and this is expected to rapidly overcome any incoherent synchrotron radiation from the start-up region in the first undulator module. In view of this, the PUFFIN results converge rapidly to that found by MINERVA and to the measured linewidths after the second undulator odule. Agreement between the simulations and the measured linewidth is within about 35% after 15 m. As shown in the figure, the predicted linewidths are in substantial agreement with the experimental measurements, and good agreement between the codes is found over the entire range of integration through the saturated regime.

## IV. SHORT PULSE SIMULATIONS

In this section, we treat the case of a short electron bunch and the SASE that is produced using the configuration of the SPARC FEL in which we include only the six undulators used in the experiment for a total length of about 15 m. In this scenario, we study the SASE interaction for a variety of electron bunch lengths in order to determine at what point the SVEA model can accurately simulate the physics of the interaction. In particular, we consider three cases corresponding to rms electron bunch lengths/durations charges of 5.59 µm/18.3 fs, 11.2 µm/36.7 fs, and 16.8 µm/55.0 fs with bunch charges of 2.9 pC, 5.8 pC, and 8.7 pC respectively. The cooperation length is approximately 21 µm for each of these cases. Hence, the three bunch lengths under consideration are all shorter than the separation distance between SASE spikes, and we expect to see single pulses produced in each of these cases rather than the series of spikes usually associated with SASE.

In many FEL simulations using the SVEA, the number of temporal slices used does not always "fill" the entire time window. This means that if the time window is $N$ wavelengths long, then fewer than $N$ temporal slices might be used. However, since we are interested in studying short pulse interactions, it is important to use contiguous temporal slices in all the MINERVA simulations so that the number of temporal slices corresponds to the number of wavelengths in the time window. On the other hand, PiC models implicitly simulate a continuous optical and electron pulse.

As there will be substantial slippage between the optical pulse and the electron bunch over the course of six undulators, it is important to choose a temporal simulation window in MINERVA which is of sufficiently broad duration that no optical power exits it. In order to ensure that this doesn't happen, we use time windows corresponding to 1000 wavelengths (1.64 ps), 2000 wavelengths (3.28 ps), and 3000 wavelengths (4.92 ps) respectively for the three bunch lengths under consideration. This corresponds to 1000, 2000, and 3000 temporal slices in the MINERVA simulations for these cases.

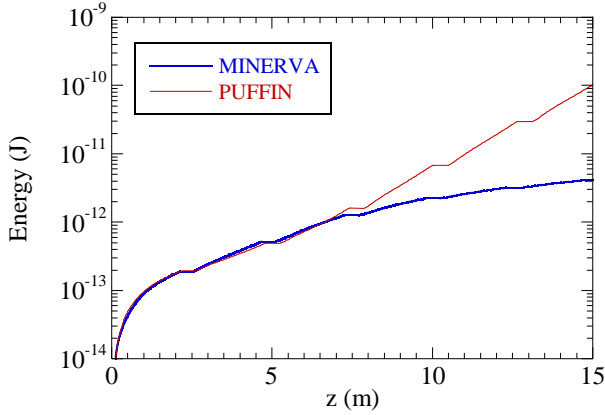

Fig. 4: Energy comparison for an rms electron bunch length of 5.59 μm.

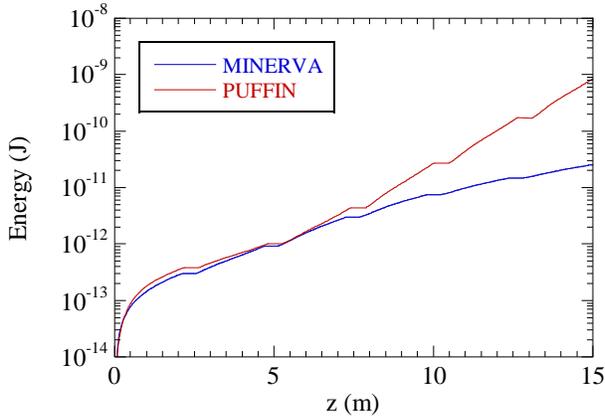

Fig. 5: Energy comparison for an rms electron bunch length 11.2 μm.

Comparisons of the evolution of the average pulse energies along the undulator lattice between PUFFIN and MINERVA for the three bunch lengths under consideration are shown in Figs. 4 – 6. As is evident in Figs. 4 and 5, while PUFFIN and MINERVA are in substantial agreement at the start up and through the 3rd undulator for the two shorter bunch lengths, they diverge significantly thereafter. We attribute this divergence either, or both, to (1) the breakdown of the SVEA, and (2) to the strong coherent synchrotron radiation (CSR) in the short bunches which is captured in PUFFIN but not MINERVA. However, as shown in Fig. 6, the two codes are in good agreement for the longest electron bunch which we attribute to the decreased importance of CSR at this bunch length. It is important to remark, though, that this bunch length is still shorter than the separation between SASE spikes.

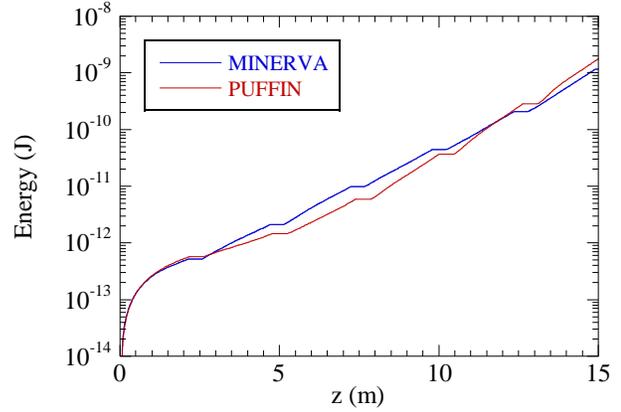

Fig. 6: Energy comparison for an rms electron bunch length of 16.8 μm.

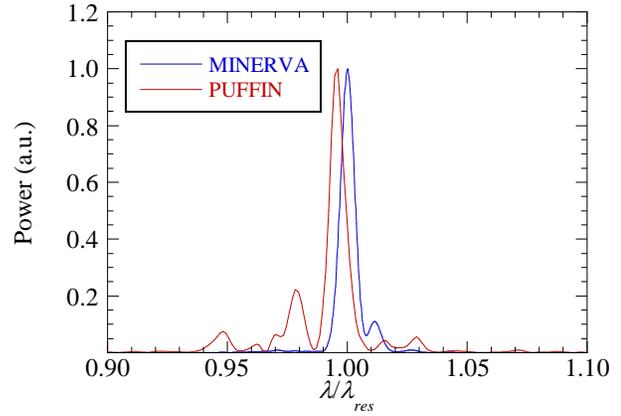

Fig. 7: Comparative spectra after the 1st undulator.

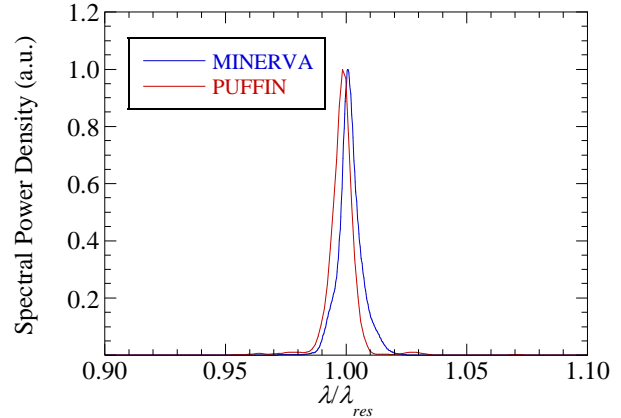

Fig. 8: Comparative spectra after the 3rd undulator.

We now consider the properties of the emission for the longest bunch length of 16.8 μm. Comparisons of the spectra obtained from the two codes after the 1st, 3rd, and 6th undulators are shown in Figs. 7 – 9 where we plot the normalized spectral power density vs wavelength (normalized to the resonant wavelength). Note that these comparisons are made using a single noise seed in each code.

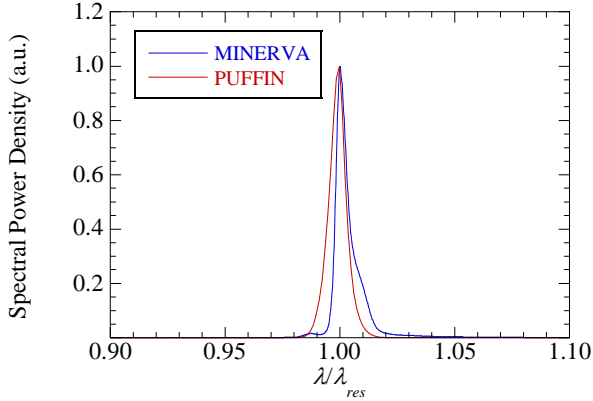

Fig. 9: Comparative spectra after the 6th undulator.

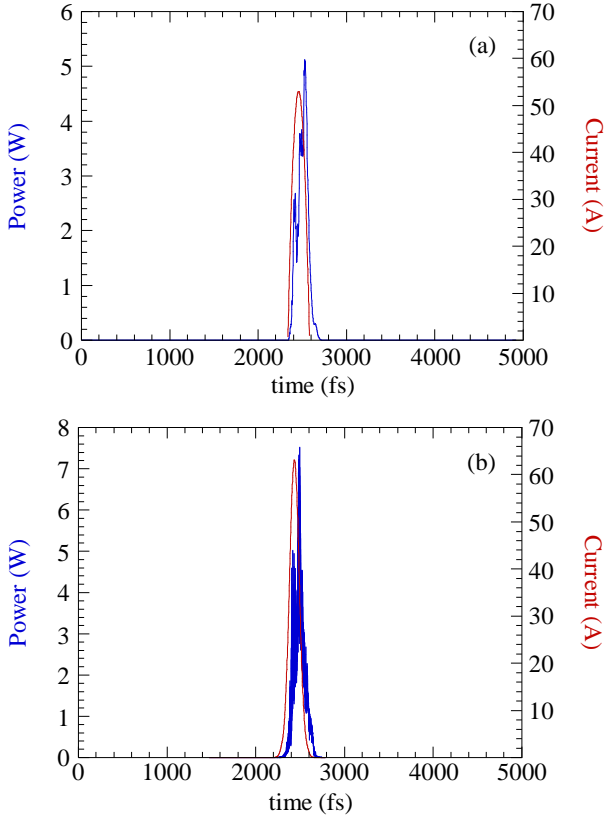

Fig. 10: Temporal pulse and current profiles after the 1st undulator from MINERVA (a) and PUFFIN (b).

As shown in Fig. 7, the peak in the spectra obtained from the two codes is shifted in wavelength by only about one part in $5 \times 10^3$ after the 1st undulator but shrinks to about one part in $10^4$ after the 6th undulator. We attribute this to the larger influence of CSR at the early stage of the interaction which is increasingly overwhelmed by the exponentiation over longer distances. However, PUFFIN exhibits a much broader bandwidth than MINERVA because PUFFIN does not make the resonant wave approximation and implicitly includes a complete spontaneous excitation spectrum. Exponential growth begins in the 2nd undulator and the resonant wavelength has come to dominate the spectrum by the end of the 3rd undulator, as shown in Fig. 8 where the spectra produced by the two codes are very close and PUFFIN now shows a peak near the resonant wavelength. The comparison of the spectra after the 6th undulator follows that after the 3rd undulator as shown in Fig. 9.

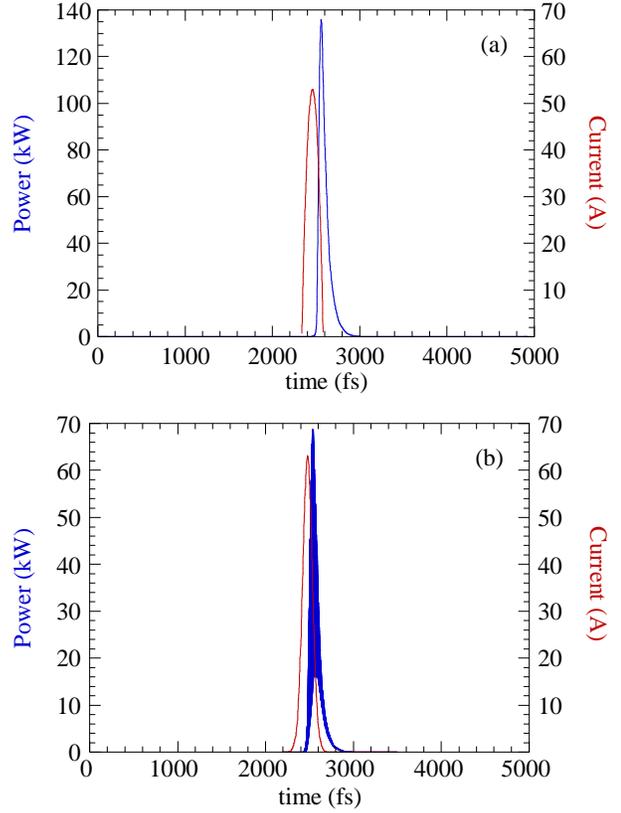

Fig. 11: Temporal pulse and current profiles after the 6th undulator from MINERVA (a) and PUFFIN (b).

The temporal pulse shapes after the 1st and 6th undulators are shown in Figs. 10 and 11 where the left axis corresponds to the radiation pulse while the right axis describes the electron bunch. In addition, the horizontal axis corresponds to the complete time window used in the MINERVA simulations. The horizontal axis for the PUFFIN results are mapped onto the same time axis as found in MINERVA. It should also be remarked that the pulse shape found in MINERVA is calculated using the Poynting flux averaged over the resonant period while that for PUFFIN is calculated using the instantaneous Poynting flux.

The relation between the pulse and current profiles after the 1st undulator are shown in Fig. 10. Note here that these results are for a single noise seed in each code so that the absolute power levels need not agree as closely as the average energy vs $z$ plots would indicate. It is clear from the figure that both codes shown similar levels of slippage. Similar observations can be made after the 6th undulator.

It should be remarked that such short electron bunch operation produces a single optical pulse rather than the usual sequence of spikes associated with SASE.

## V. SUMMARY AND CONCLUSION

In this paper, we simulated the FEL interaction using ultra-short electron bunches and described a comparison between FEL models that rely on the SVEA formulation, which is represented by the MINERVA code, and on the complete Maxwell's equations, which is represented by the PiC formulation code PUFFIN. The models have been applied to the SPARC SASE FEL at ENEA Frascati. Good agreement has been found both between the two models and between the models and the experiment, thereby validating both formulations.

This is significant because the underlying simulation codes have virtually no elements in common, and we can conclude from this that they both faithfully describe the physics underlying FELs. In particular, the agreement between the codes and the experimental measurements regarding the start-up regime in the SPARC FEL validates the different particle loading algorithms in both codes.

Limitations of the SVEA models derive from the fast time scale average and from the inability to treat CSR from first principles. These limitations are not present in PiC codes. To this end, we have also compared the codes for a set of parameters similar to that from the SPARC experiment except that we have dealt with ultra-short electron bunches. We considered three cases corresponding to bunch lengths that are 5.59 μm, 11.2 μm, and 16.8 μm in extent which are shorter than the cooperation length of 21 μm. Hence, each of these bunch lengths are shorter than the separation of SASE spikes and we found that the spontaneously generated radiation consists of a single pulse rather than the usual spiky SASE radiation [1,2]. We find that while the MINERVA results differ substantially from the PUFFIN results for the two shortest bunch lengths, the two codes are in substantial agreement for the longest bunch which is comparable to, but slightly shorter than, the cooperation length. Since the longest bunch still represents a bunch that is shorter than the SASE spike separation, we conclude that while a PiC code like PUFFIN is the more general of the two codes, the SVEA model may still have utility in modeling some short bunch configurations.

For example, the bunch duration/length for the LCLS [13], which is an x-ray SASE FEL operating at a wavelength of 1.5 Å, is about 83 fs/25 μm and the cooperation length is about 20 nm. The agreement between PUFFIN and MINERVA using the SPARC parameters for a bunch length comparable to the cooperation length suggests that the SVEA is able to faithfully simulate rms electron bunch lengths/durations of about 67 as/20 nm in an LCLS-like device. However, simulation of such short bunches in an LCLS-like device may require the development of a self-consistent model for CSR in the SVEA codes. This may be important because such short bunch operation that results in a single radiation pulse is characterized by near-transform limited pulses, but the computational requirements needed to simulate such an FEL with a PiC code such as PUFFIN may be prohibitively large.

Areas where PiC simulations may be required are where the electron beam and/or radiation envelope properties change significantly in one radiation period. An example of such fast changes is in the generation of short, high-power pulses – see e.g. [14,15]. Previous results of high-power propagation in 1D simulations [16] have already illustrated significant differences between the SVEA and PiC models and this area will be the subject of further research.

## ACKNOWLEDGEMENTS

The research used resources of the Argonne Leadership Computing Facility, which is a DOE Office of Science User Facility supported under contract DE-AC02-06CH11357. We also thank the University of New Mexico Center for Advanced Research Computing, supported in part by the National Science Foundation, for providing high performance computing resources used for this work.

Funding is also acknowledged via the following grants: Science and Technology Facilities Council (Agreement Number 4163192 Release #3); ARCHIE-WeSt HPC, EPSRC grant EP/K000586/1; John von Neumann Institute for Computing (NIC) on JUROPA at Jülich Supercomputing Centre (JSC), project HHH20.

The authors acknowledge helpful discussions with L. Giannessi.